\documentclass[11pt,titlepage]{report}
\usepackage{Preamble}
\usepackage{program}

\begin{document}

\begin{titlepage}
    \newgeometry{margin=3cm}
	\centering
    \includegraphics[width=0.5\linewidth]{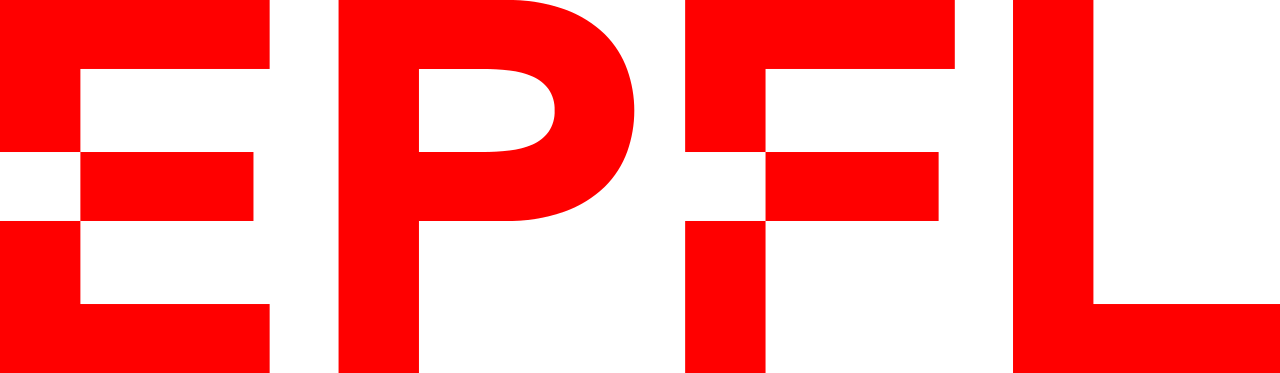}\\[0.25cm]
    \textsc{\LARGE École Polytechnique Fédérale de Lausanne}\\ \vspace{\fill}
    \textbf{\textsc{\fontsize{50}{50}\selectfont Report}}\\\vspace{1.5in}
    \textsc{\LARGE  Exploring ensembles and uncertainty minimization in denoising networks}\\
    \vspace{\fill}		
	\textsc{\LARGE Semester project \vspace{10pt}\\Image and Visual Representation Lab (IVRL)}\\[0.4cm]
	\rule{\linewidth}{0.2 mm} \\[0.5 cm]
	Student: Xiaoqi MA\\Supervisior: Majed El Helou\\Professor: Sabine S\"usstrunk\\[2cm] \today
\end{titlepage}
\restoregeometry

\thispagestyle{numberonly}
\pagenumbering{roman}
\chapter*{Abstract}
The development of neural networks has greatly improved the performance in various computer vision tasks. In the filed of image denoising, convolutional neural network based methods such as DnCNN\cite{zhang2017beyond} break through the limits of classical methods, achieving better quantitative results. However, the epistemic uncertainty existing in neural networks limits further improvements in their performance over denoising tasks. Therefore, we develop and study different solutions to minimize uncertainty and further improve the removal of noise.

From the perspective of ensemble learning, we implement manipulations to noisy images from the point of view of spatial and frequency domains and then denoise them using pre-trained denoising networks. We propose a fusion model consisting of two attention modules, which focus on assigning the proper weights to pixels and channels. The experimental results show that our model achieves better performance on top of the baseline of regular pre-trained denoising networks. 

\textbf{Keywords: Deep denoisers, epistemic uncertainty, ensemble methods, neural attention.}

\pagenumbering{arabic}
\chapter{Introduction}
Image denoising is a fundamental low-level computer vision task, with considerable attention in recent years especially with the development of neural network methods. The images captured in the real world are always inevitably degraded by noise, which will cause the quality of images to decrease, also affecting the performance of other subsequent image processing or computer vision tasks such as classification. In order to reduce the noise and improve image quality, some effective methods have been proposed in the field of image denoising. For instance, classical methods such as BM3D\cite{BM3D}, which is based on enhanced sparse representation are aimed at removing noise at certain noise levels. Neural networks provide a novel alternative for better removing noise from images. Currently, denoising networks\cite{zhang2017beyond,tai2017memnet,anwar2019real,el2020blind,elhelou2020stochastic} achieve state-of-the-art performance under the blind setting of noise levels.

The goal of our project is to explore the uncertainty existing in the denoising networks. These convolutional neural networks are used for removing Additive White Gaussian Noise. Considering the epistemic uncertainty of the networks themselves, we intend to minimize such uncertainty from the angle of self-ensemble methods. Therefore, establishing suitable and reasonable branches for ensembling is an important cornerstone. From the perspective of the spatial (pixel) domain and frequency domain, we proposed two kinds of manipulation methods: Rotation and Flip, and DCT masking to form the ensemble branches. With different manipulated noisy images, the pre-trained denoising networks can produce the corresponding manipulated denoised images. We then use our fusion method to combine these images in order to get better performance.

The fusion model is established with dual attention mechanism: spatial position attention module and channel attention module. These two modules are parallel processing the manipulated images at first and make fusion of results to extract the final images. Spatial attention module is used for producing weight maps for pixels and channel attention module focuses on weighing different manipulated images. The experimental results show that our model makes good progress at image denoising. Besides, we also tried other ideas like training the denoising network with auxiliary loss, but the results reflect that it still needs more improvements.

In summary, our project aims to analyze and reduce the uncertainty in denoising networks. We hope to find reasonable and effective methods to minimize the uncertainty with small computational cost. Through comparative analysis of experimental results, we find that the solutions we proposed have further reduced the uncertainty in denoising networks to varying degrees.

\chapter{Literature review}
\section{Image Denoising}
\subsection{Classical denoising Methods}
Most classical denoising methods take advantage of the correlation between pixels to remove the noise which has a low correlation with pixels of clean images. The spatial domain filtering consists of linear filters such as mean filtering\cite{zhang2009impulse} and non-linear filters like median filtering\cite{justusson1981median}. It is easy to implement spatial filtering, but it removes noise at cost of blurring. In addition, lots of variational denoising methods are used to remove noise by minimizing the energy function with image priors. Total Variation (TV)\cite{vogel1996iterative} methods achieve great performance at image denoising and decrease the effect of smoothness. K-singular value decomposition (KSVD)\cite{lebrun2012implementation} takes advantages of dictionary learning and sparse coding to represent the image features and thus removes the noise with less similarities. Otherwise, low-rank factorization is proposed by decomposing into low rank matrices and then used to remove the noise\cite{jha2010denoising}.

Gradually, the transform methods are developed at image denoising such as Fourier transform, discrete cosine transform, wavelet transform. Principal Component Analysis (PCA)\cite{muresan2003adaptive} is used as a transform tool by local pixel grouping to remove the noise. Block-matching and 3D filtering (BM3D) groups similar 2D image blocks into 3D groups and use collaborative filtering to extract the true images without noise. These classic denoisers which are aimed at Gaussian denoising are mostly target at denoising with certain noise levels.

\subsection{Deep Denoisng Neural Networks }
The Neural Networks achieve outstanding performance in image processing tasks such as image classification, segmentation and restoration. As the simplest model, MLP-based models work efficiently by a feed-forward deep network. DnCNN is proposed to remove the additive white Gaussian noise (AWGN) of images by combining the convolutional layers, batch normalization and activation layers. With the notion of residual learning, DnCNN learns the noise in the image instead of image itself. Under the blind settings, DnCNN trains the denoising network while varying the training noise levels and can predict any noise level between $(0,55)$. Otherwise, some denoising networks such as MemNet\cite{tai2017memnet}, RIDNet\cite{anwar2019real} are also designed for Gaussian denoising under blind settings. From the perspective of frequency domain, the Stochastic Frequency Masking (SFM)\cite{elhelou2020stochastic} method leverages the conditional learning in the training phase and improves the high noise level results in image denoising tasks.

\section{Uncertainty in Deep Neural Networks}
According to the research\cite{kendall2017uncertainties} about uncertainties in deep learning for computer learning, there exist two kinds of uncertainties in Deep Neural Networks (DNNs): one is the Aleatoric Uncertainty (statistical uncertainty), which is caused by data error and is unavoidable even if more data are provided. The other is Epistemic Uncertainty (systematic uncertainty), which measures the uncertainty in the model. The Epistemic Uncertainty is caused by the limited data and knowledge, so it can be decreased by obtaining enough training samples and better models.

Multiple methods are proposed to deal with the Epistemic Uncertainty in the deep neural networks. Bayesian Neural Networks (BNNs)\cite{kononenko1989bayesian} provide us with a way to predict uncertainty by learning predictive distributions. It changes standard deep neural networks which aim to find the maximum likelihood estimation (MLE) to networks with posterior inference. BNNs have extensive applications like spam filtering, gene regulatory networks and many other computer vision tasks.

Although BNNs capture the model uncertainty, they are at the cost of computational cost. Monte Carlo Dropout\cite{gal2016dropout} is proposed at the training phase as a tool to approximate Bayesian inference. It is used to measure the model uncertainty and show the confidence of models.

Classical ensemble learning\cite{zhou2009ensemble} is designed to reduce the uncertainty of the model or improve the performance of the model by combining diverse models. It reduces the risk of overfitting and improves the generalization of models. There exists a multitude of applications using ensemble learning such as data fusion or confidence estimation. Some fundamental strategies of ensemble learning in classification tasks are simple voting, weighted voting, model stacking, boosting and random forests.

% Summary of datasets used in related works are listed in Table \ref{tab:dataset}.
% \input{table/dataset}

\chapter{Methods }
\section{Dual attention fusion model}
\subsection{Overall Design}
In order to analyze the epistemic uncertainty existing in the Deep Neural Networks, we propose an fusion model consisting of two main attention modules: spatial attention module and channel attention module. The key element of our method is based on ensemble of different branches manipulated from two angles: spatial domain and frequency domain. With the pre-trained denoising networks, we can get the corresponding denoised images from different manipulated noisy images, and then make the manipulated pre-denoised images become the input of our fusion models. In the spatial domain, we consider the flip and rotation operations as manipulation modes, and in the frequency domain we consider the DCT mask as the manipulation modes.
\subsection{Manipulation methods}
\subsubsection{Flip and rotation}
\begin{figure}
    \centering
\subfloat[Removed noise of normal images\label{fig:w1}]{
    \includegraphics[width=0.3\textwidth]{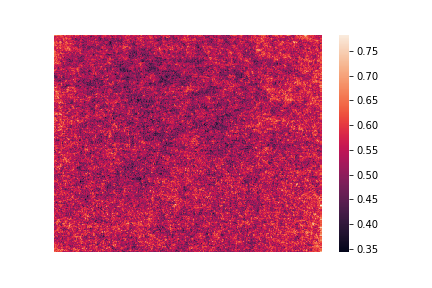}}
\subfloat[Removed noise of rotated images(90 degree)\label{fig:w2}]{
    \includegraphics[width=0.3\textwidth]{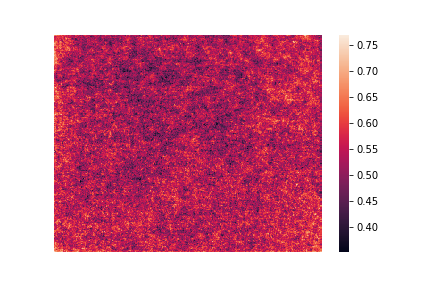}}
\subfloat[Removed noise difference between normal and rotated images\label{fig:w3}]{
    \includegraphics[width=0.3\textwidth]{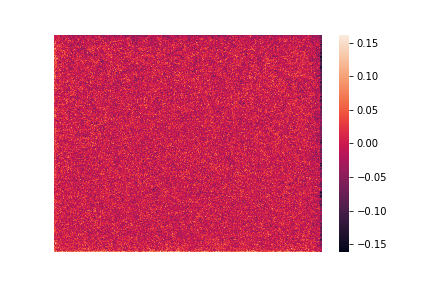}}
\caption{Average removed noise heat-map of the normal images and rotated images.}
\label{fig:heat-map}
\end{figure}

From the perspective of spatial domain, we use the operation of flip and rotation as the manipulation modes to do explore further ensemble strategy. The experimental results show that pre-trained denoisers are not invariant to the image mirroring and rotation although such data augmentations are used during the training phase. From the figure \ref{fig:heat-map}, we can find slightly different Peak Signal to Noise Ratio(PSNR) results for flipped and rotated images from denoisers.

In our experiments, we propose seven kinds of manipulation modes to transform the noisy images:
\begin{itemize}
\vspace{0pt}
    \setlength{\itemsep}{0pt}
    \setlength{\parsep}{0pt}
     \setlength{\parskip}{0pt}
    \item mode 0: initial image 
    \item mode 1: rotate \ang{90} and vertical mirroring
    \item mode 2: vertical mirroring
    \item mode 3: rotate \ang{270}
    \item mode 4: rotate \ang{180} and vertical mirroring
    \item mode 5: rotate \ang{90}
    \item mode 6: rotate \ang{180}
    \item mode 7: rotate \ang{270} and vertical mirroring
\end{itemize}

\subsubsection{DCT mask}
\begin{figure}
    \centering
\subfloat[$\sigma=10$ \label{fig:q1}]{
    \includegraphics[width=0.3\textwidth]{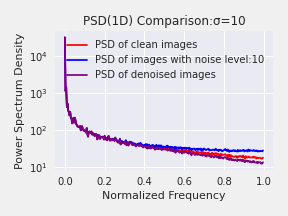}}
\subfloat[$\sigma=30$ \label{fig:q2}]{
    \includegraphics[width=0.3\textwidth]{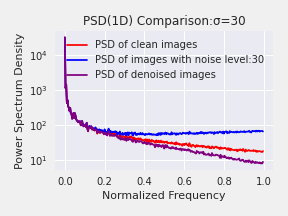}}
\subfloat[$\sigma=50$\label{fig:q3}]{
    \includegraphics[width=0.3\textwidth]{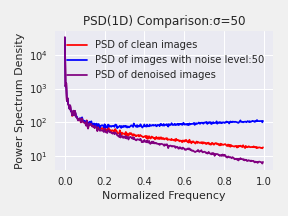}}
\caption{Power Spectrum Density (PSD) comparison for different noise levels of test data set.}
\label{fig:psd}
    \end{figure}

From the perspective of frequency domain, we use the Stochastic Frequency Masking as the manipulation modes. Firstly, the discrete cosine transform (DCT) expresses a finite sequence of data points in terms of a sum of cosine functions oscillating at different frequencies. As we all know, for the DC coefficients matrices of images, most of the energy of the images is concentrated in the low frequency domain, by contrary, the energy of image noise, details such as edges is concentrated in the high frequency domain. From Fig \ref{fig:psd}, which shows the Power Spectral Density (PSD) of clean images, noisy images and denoised images, we can find that the frequency components of noise are mostly concentrated in the high frequency domain and denoising networks mainly remove the high frequency components of the images.

After transforming the image to the frequency domain using the Discrete Cosine Transform (DCT), we delimit the quarter-annulus area by setting the values of its inner and outer radius. We implement two kinds of masking types: mask the area after a given radius; mask the area between the given inner and outer radius. The first three types are used for filtering out what corresponds to high frequencies for varying noise levels and the last two types are designed for masking the mid-to-high frequencies. The $radius_{max}$ is defined as the maximum diagonal of DC coefficient matrix of image.

\begin{figure}
    \centering
    \captionsetup{font={small}}
\subfloat[Clean images + Noisy images \label{fig:s1}]{
    \includegraphics[width=0.21\textwidth]{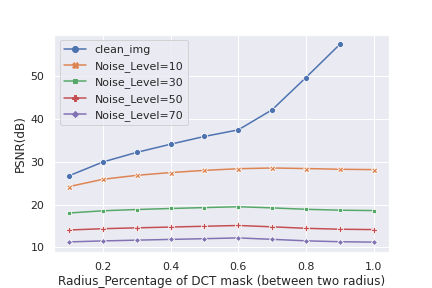}}
\subfloat[Denoised images \label{fig:s2}]{
    \includegraphics[width=0.21\textwidth]{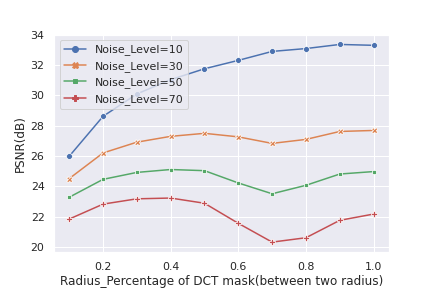}}
\subfloat[Clean images + Noisy images \label{fig:s3}]{
    \includegraphics[width=0.21\textwidth]{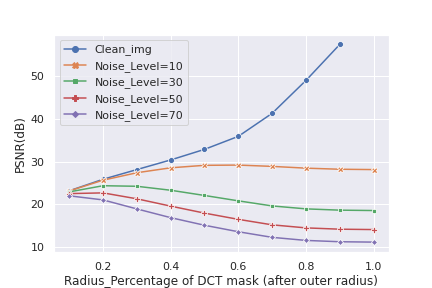}}
\subfloat[Denoised images \label{fig:s4}]{
    \includegraphics[width=0.21\textwidth]{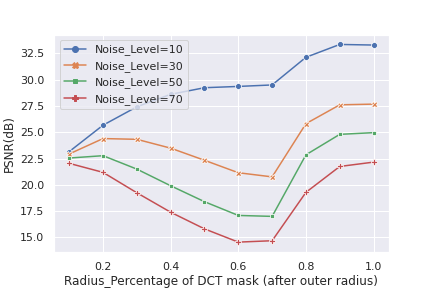}}
\caption{PSNR(dB) comparison for DCT masking for clean images, noisy images and denoised images with two DCT mask types. The Fig \ref{fig:s1} and Fig \ref{fig:s2} represent the DCT mask between given inner radius and outer radius. The Fig \ref{fig:s3} and Fig \ref{fig:s4} represent the DCT mask after one given radius.}
\label{fig:psnr——dct}
    \end{figure}

The Fig \ref{fig:psnr——dct} shows the PSNR change along with the different mask radius values for clean images, noisy images and denoised images. We can find that the different mask strategies have different effect on the noise removal.

\begin{itemize}
    \setlength{\itemsep}{0pt}
    \setlength{\parsep}{0pt}
    \setlength{\parskip}{0pt}
    \item mode 8: mask the area after $radius_{max} * 0.1$
    \item mode 9: mask the area after $radius_{max} * 0.3$
    \item mode 10: mask the area after $radius_{max} * 0.5$
    \item mode 11: mask the area between $radius_{max} * 0.4 $ and $radius_{max} * 0.5$
    \item mode 12 :mask the area between $radius_{max} * 0.5 $ and $radius_{max} * 0.9$
\end{itemize}

\subsection{Fusion Models}
After implementing the image manipulations to noisy images by all the manipulation types shown above, we use the pre-trained denoisers to produce corresponding pre-denoised images. Our fusion model is based on spatial position attention mechanism and channel attention mechanism.

\begin{figure*}
\centering
\includegraphics[width=\linewidth]{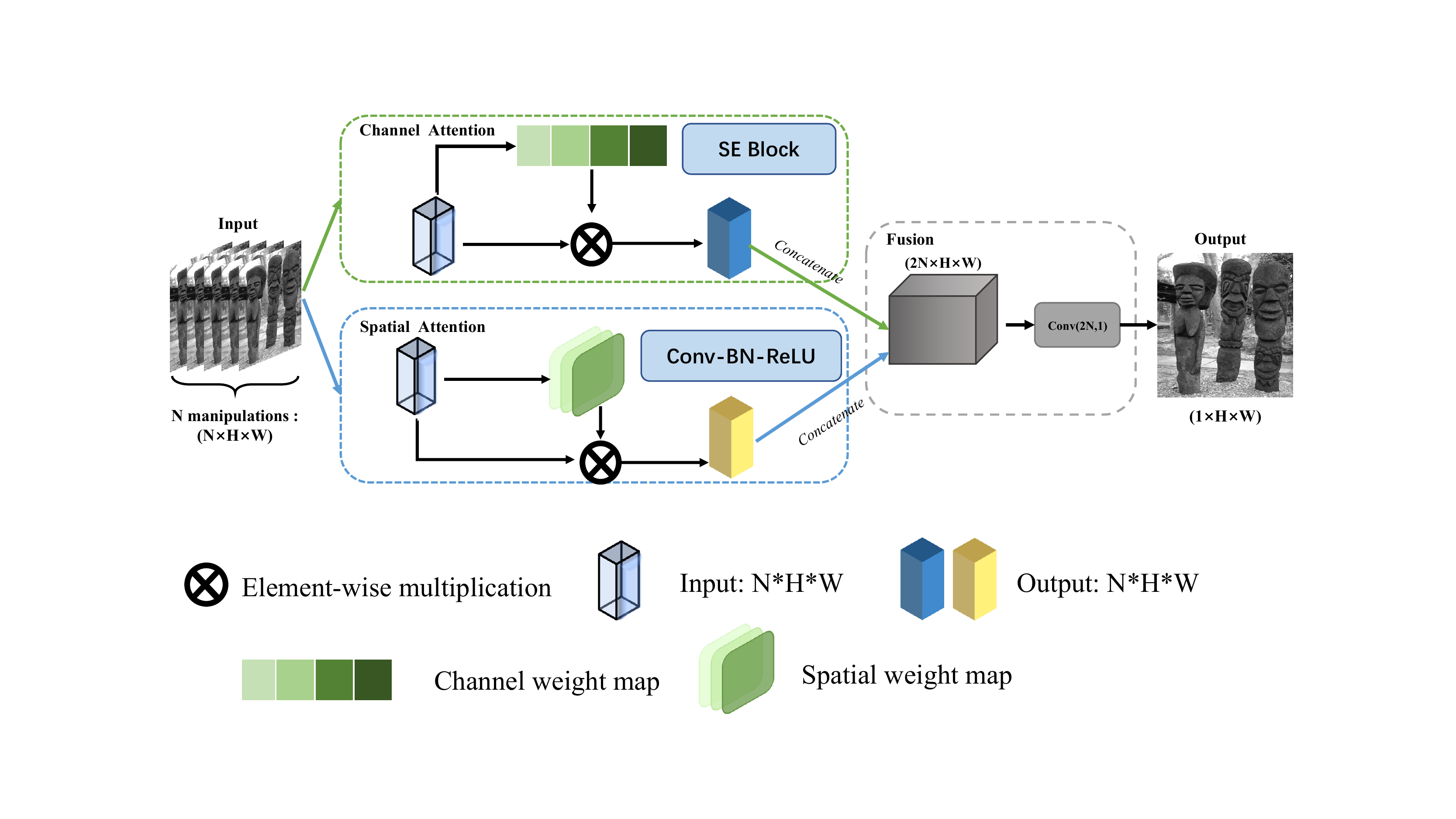}
\caption{Fusion model architecture}
\label{fig:net_structure}
\end{figure*}

The spatial attention module is designed for producing a spatial weight map sharing the same size as image itself, and then the weight map conducts element-wise multiplication with the corresponding input manipulated images. At this module, we let the N kinds of manipulated images concatenate at channel dimension of the input, then these manipulated images use convolution layers, batch normalization and activation functions to extract N corresponding weight maps, and we implement the Softmax function at channel dimension of maps and do element-wise multiplication at last. On the other hand, the channel attention module focuses on assigning weights to each channel (different manipulated images) by Squeeze-and-Excitation (SE) block. It also utilizes Softmax function to guarantee weights assigned to channels sum to 1. These two modules are parallel processing the input manipulated images and at last, we fuse results from two modules by convolution layer. The Network architecture is shown on Fig \ref{fig:net_structure}. 

\section{Retrain DnCNN with Auxiliary Loss}
In order to conduct in-depth analysis of epistemic uncertainty of denoising networks, we introduce an error estimator to estimate the $L_{1}$ or $L_{2}$ error of the denoisers. In other words, the target of the error estimator is the  the $L_{1}$ or $L_{2}$ error of images denoised by denoisers and clean images. The input images of error estimator are also noisy images. The network structure of error estimator is Full Convolution Network including Convolution layers, Batch Normalization and activation layers. 

\begin{figure*}
\centering
\includegraphics[width=\linewidth]{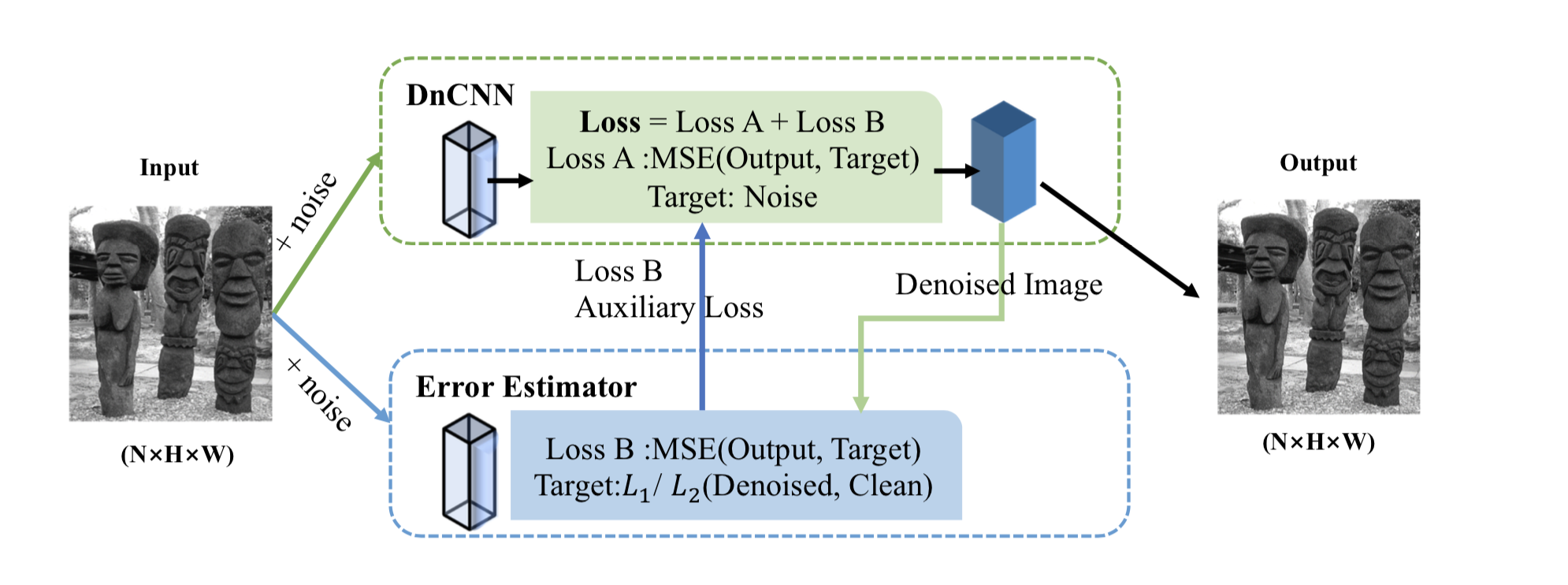}
\caption{Train DnCNN with auxiliary loss }
\label{fig:error_est}
\end{figure*}

With the usage of error estimator, we can retrain the denoiser (DnCNN) with auxiliary loss, which is produced by the error estimator. First, we use the noisy images as input to denoiser and error estimator, the denoiser will output the denoised images and give it to error estimator to calculate the target value of error estimator. Then error estimator transmits its loss to denoiser as auxiliary loss of denoiser. The denoiser  trains the network with the origin loss plus auxiliary loss. The denoiser and error estimator alternately train their own network within one batch. At last we can compare the results of regular DnCNN and DnCNN with auxiliary loss. The training progress is shown on Fig \ref{fig:error_est}

In addition, we also try to implement another auxiliary task: DnCNN which learns images. We have known the regular DnCNN learns the noise by residual learning, so we train the DnCNN with the target of images. At such circumstances, the DnCNN with target of images transmits its loss as auxiliary loss to DnCNN with target of noise (regular one).

\chapter{Results \&Discussion}
\section{Dual Attention Fusion Model}
\subsection{Experimental Parameters}
Firstly, we declare the experimental parameters about pre-traind denoisers:
\begin{itemize}
    \setlength{\itemsep}{0pt}
    \setlength{\parsep}{0pt}
    \setlength{\parskip}{0pt}
    \item The training data-set of pre-trained denoisers (DnCNN, MemNet, RIDNet) is BSD400 and the data augmentations are flip, rotation and scale resize. Each image is split into patches of patch size $64\times 64$ with stride 32. 
    \item The optimizer of denoisers is ADAM and the criterion is MSE. The initial learning rate is $1e-3$ and declines with increasing epochs, the number of epochs is 50. 
    \item The denoisers don't clip the training noisy images to [0,1], but clip the test noisy images to [0,1] before flowing into denoisers.
    \item Denoiser training noise level range: [0,55] (i.e. the denoiser was trained with blind noise levels sampled randomly from [0,55]).
\end{itemize}

Then, we declare the experimental parameters about our fusion model:
\begin{itemize}
    \setlength{\itemsep}{0pt}
    \setlength{\parsep}{0pt}
    \setlength{\parskip}{0pt}
   \item Training dataset: 200 pictures which come from test set of BSD500 and then denoised by pre-trained denoisers. Test dataset: 100 pictures which come from val set of BSD500 and then denoised by pre-trained denoisers.
   \item The images are grayscale images, because RGB or generally multispectral images have strong correlation across channels\cite{deblur} that eases the denoising and can bias the experimental results and conclusions.
    \item All the images are split into patches of patch size $50\times 50$ with stride 50. \item The optimizer is ADAM and the criterion is MSE, the number of epochs is 100. The initial learning rate is 0.01 and after 50 epochs, it declines by factor 0.6 with step size 30. 
    \item The fusion model is trained for special noise level 10, 20, 30, 40, 50.
\end{itemize}
\subsection{Results}
\begin{figure}
    \centering
\subfloat[Noise free\label{fig:t1}]{
    \includegraphics[width=0.25\textwidth]{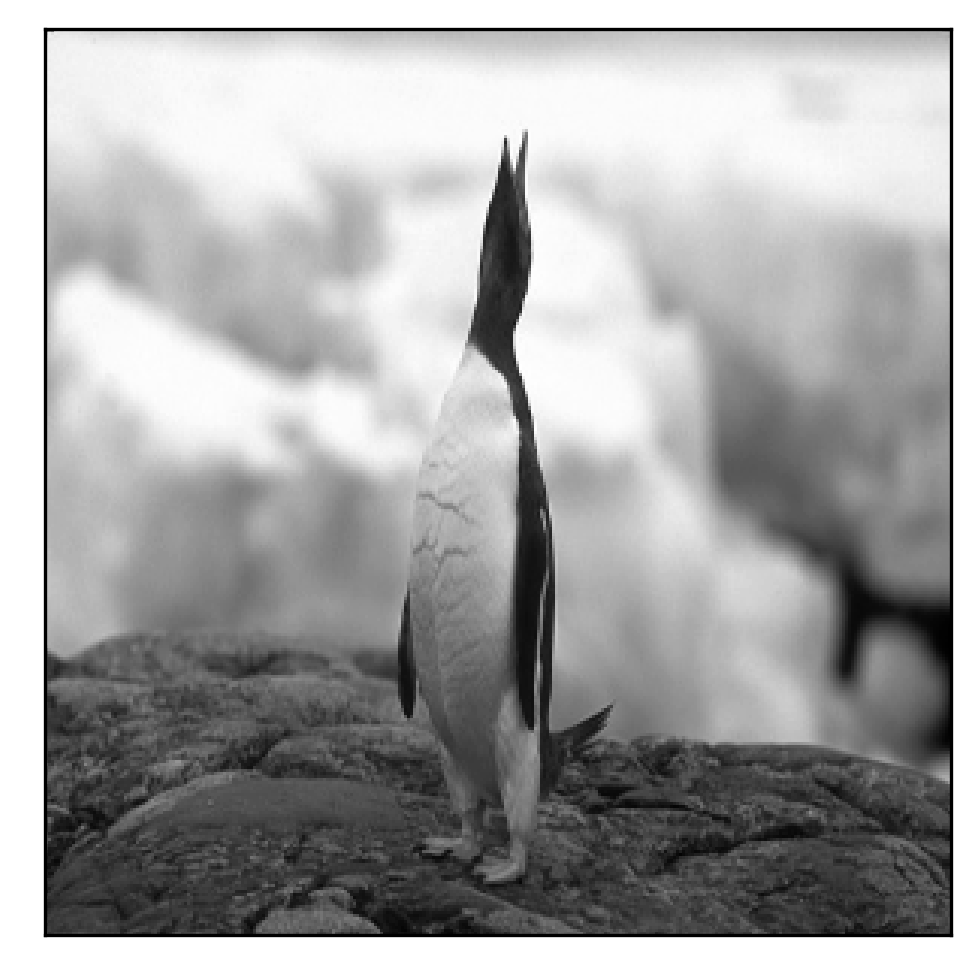}}
\subfloat[DnCNN\label{fig:t2}]{
    \includegraphics[width=0.25\textwidth]{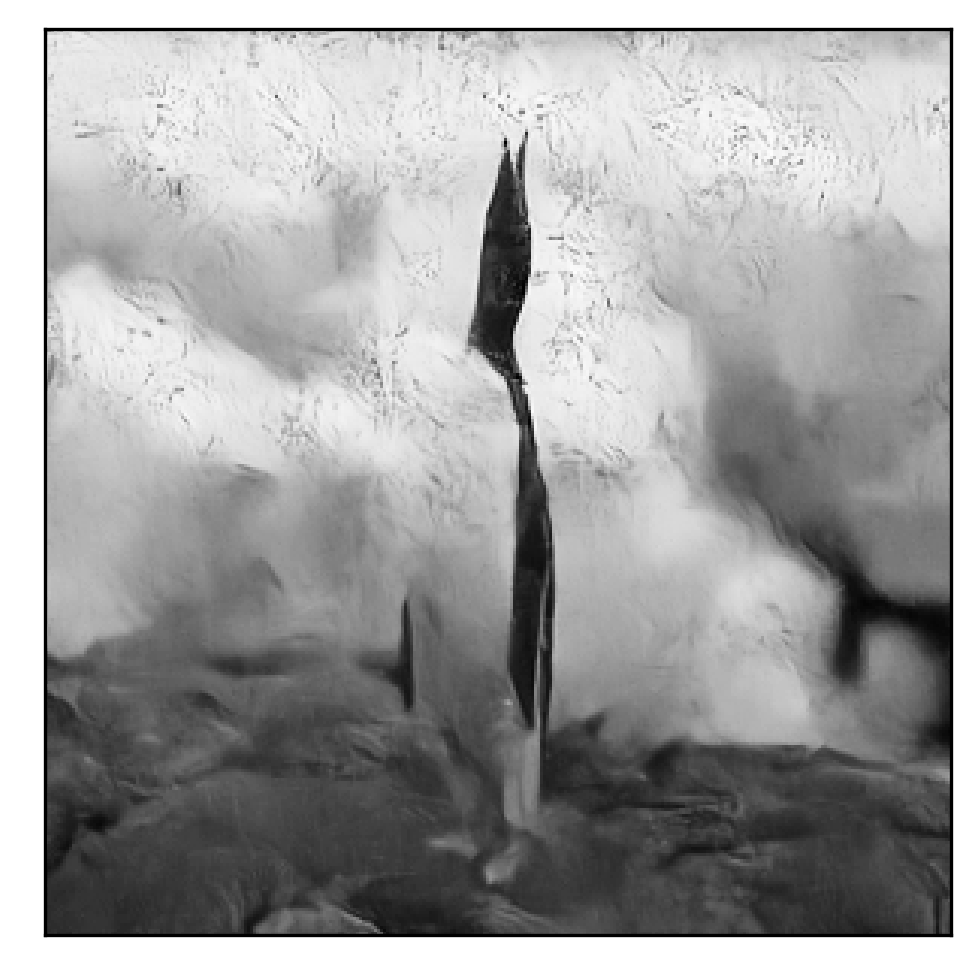}}
\subfloat[MemNet\label{fig:t3}]{
    \includegraphics[width=0.25\textwidth]{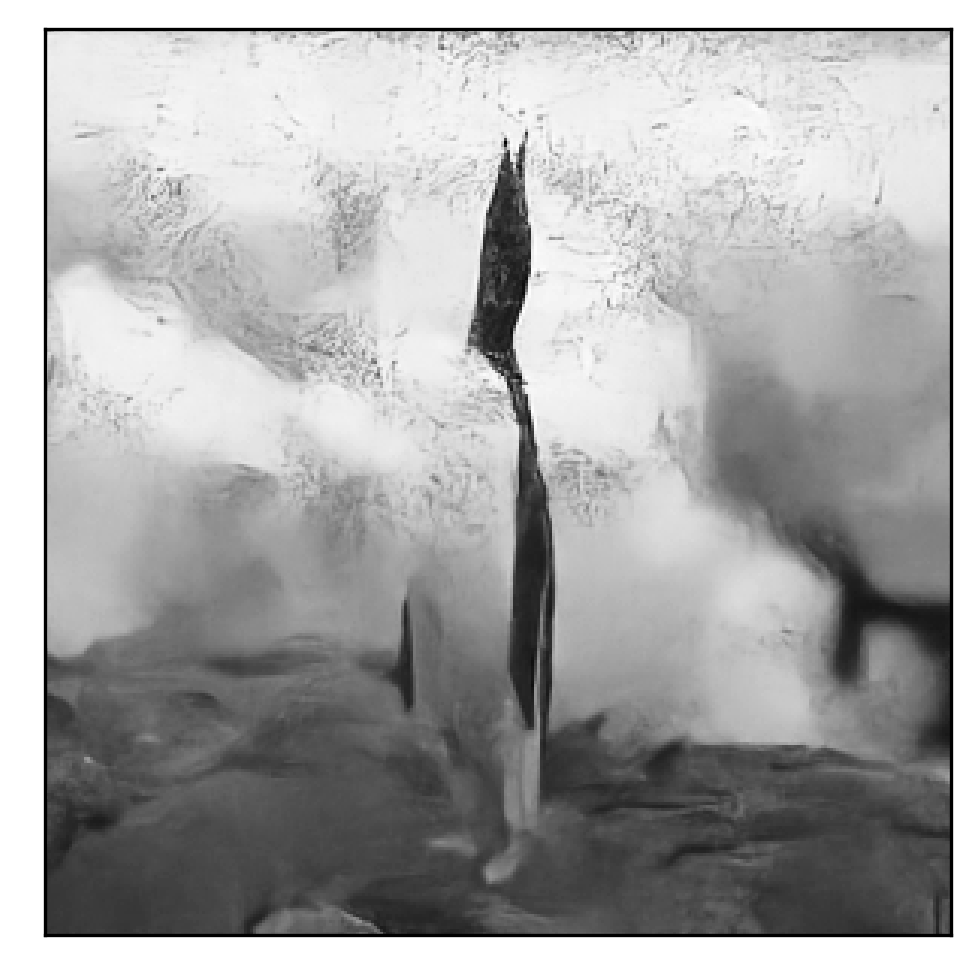}}
\subfloat[RIDNet\label{fig:t4}]{
    \includegraphics[width=0.25\textwidth]{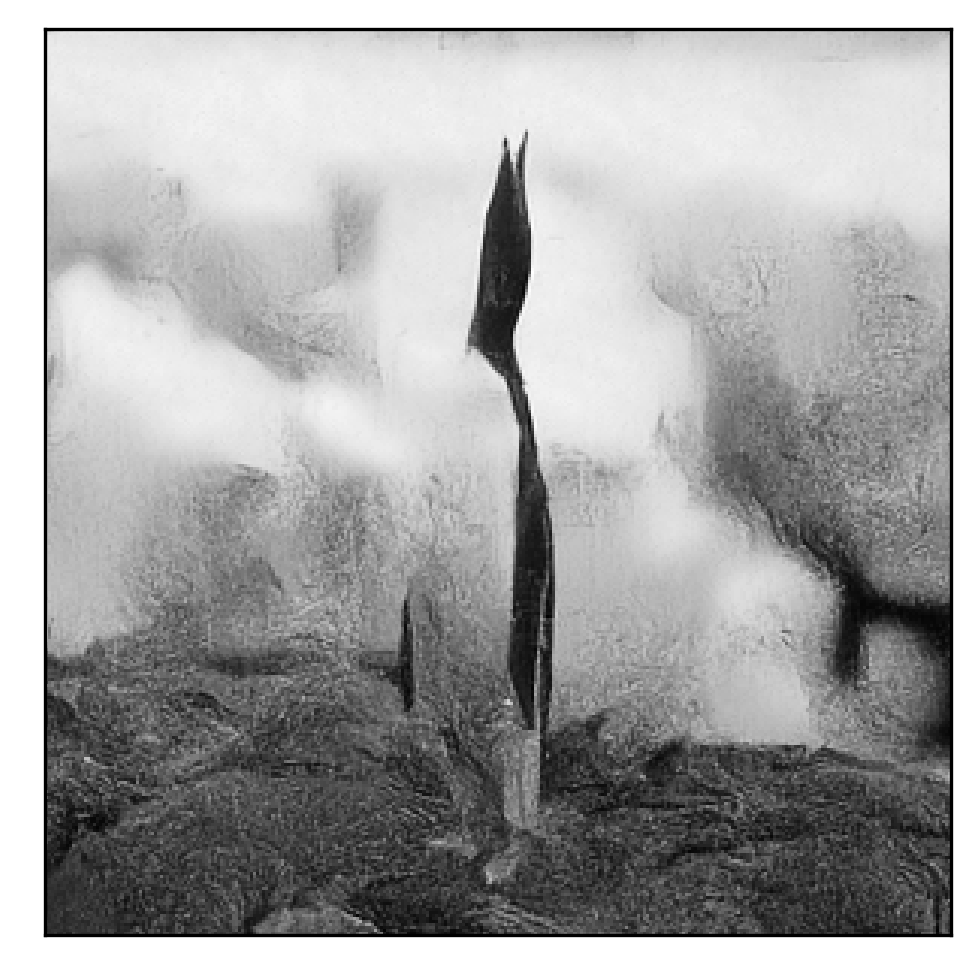}}
    
\subfloat[Noisy ($\sigma=50$) \label{fig:y1}]{
    \includegraphics[width=0.25\textwidth]{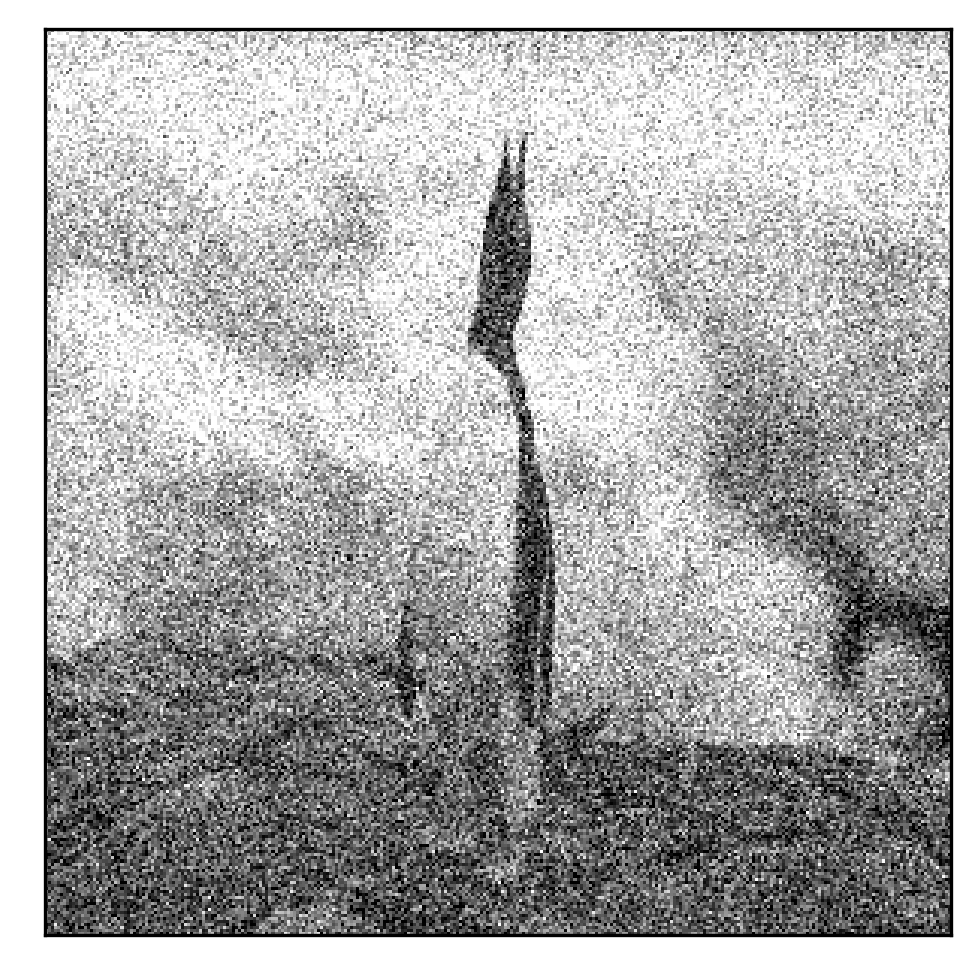}}
\subfloat[DnCNN + Fusion\label{fig:y2}]{
    \includegraphics[width=0.25\textwidth]{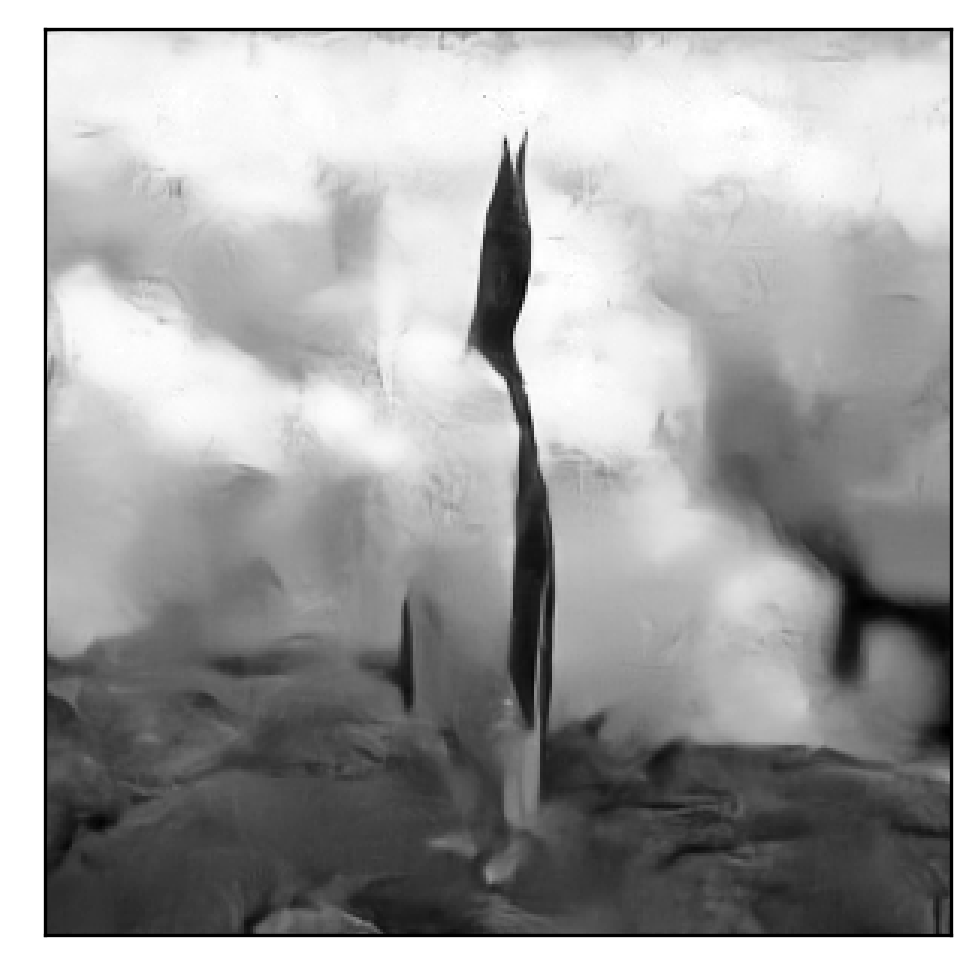}}
\subfloat[MemNet + Fusion\label{fig:y3}]{
    \includegraphics[width=0.25\textwidth]{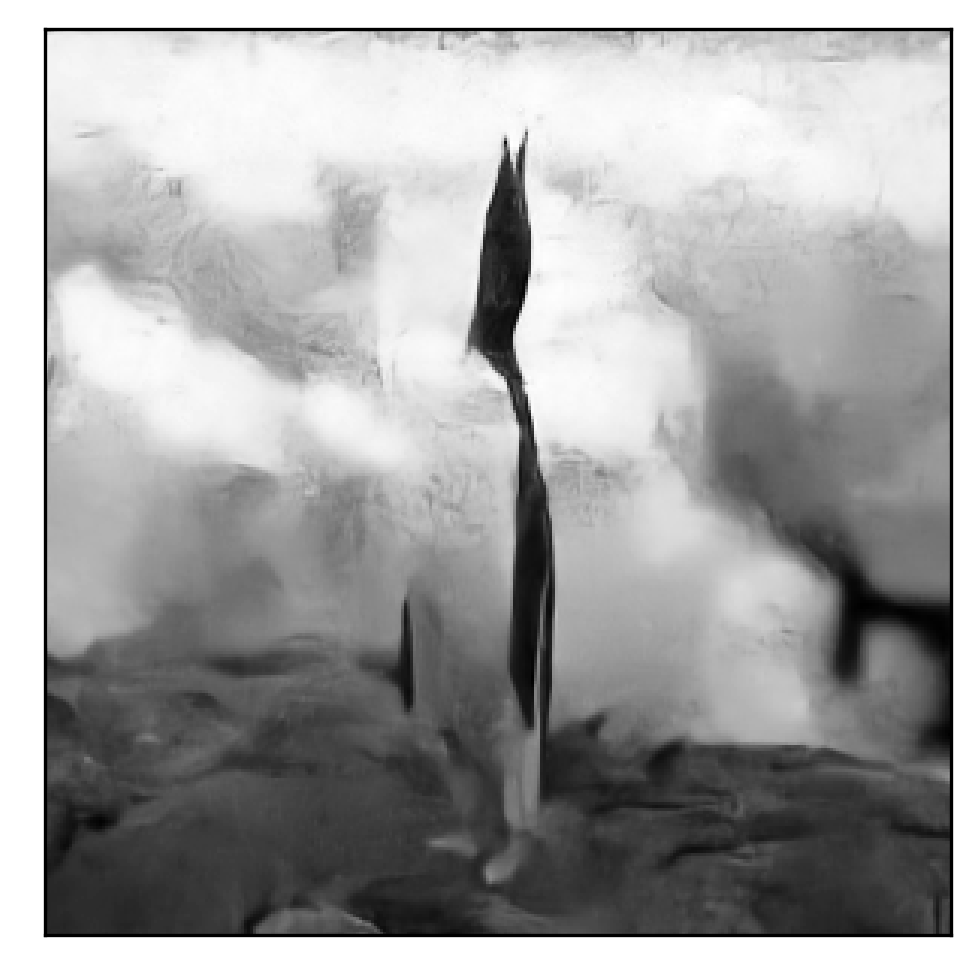}}
\subfloat[RIDNet+ Fusion\label{fig:y4}]{
    \includegraphics[width=0.25\textwidth]{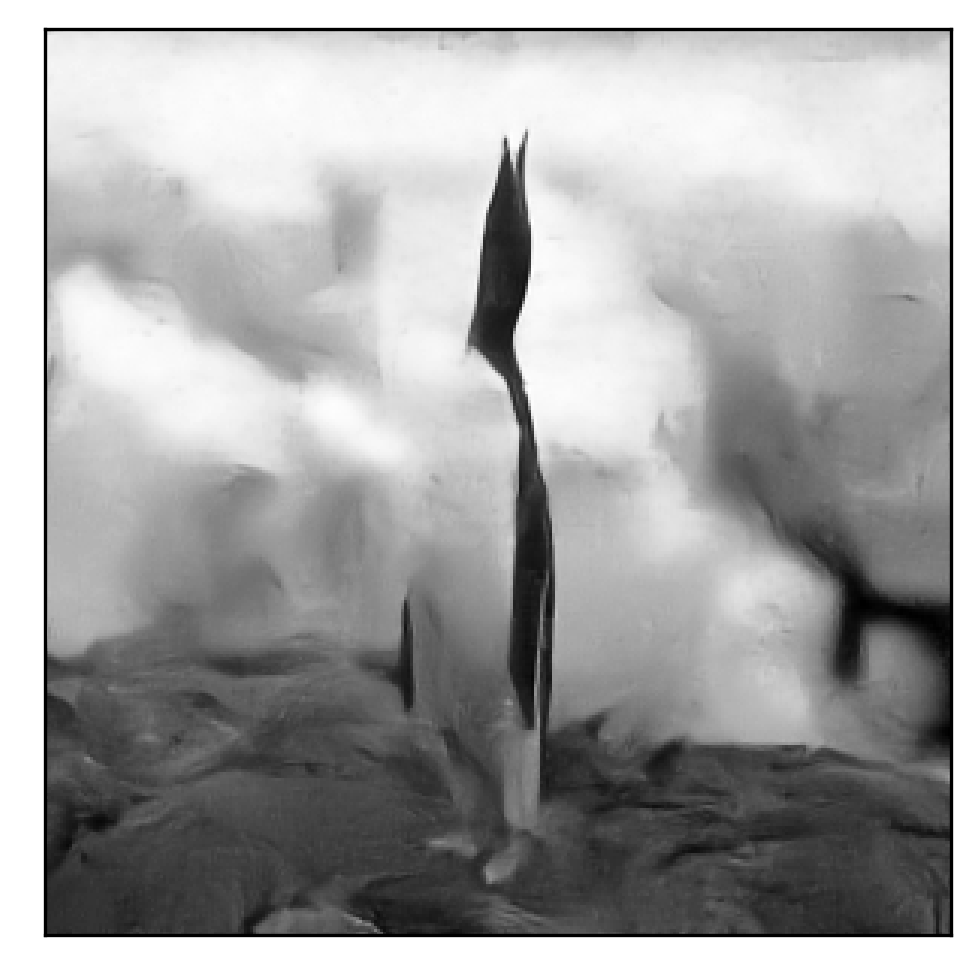}}
\caption{Sample visual denoising results of different baselines (top row), and the corresponding results with our fusion method (bottom row), for the noise level $\sigma=50$.}
\label{fig:dncnn-al}
    \end{figure}

We train our fusion model based on three kinds of pre-trained denoisers: DnCNN, MemNet and RIDNet. In order to compare effect of different attention modules, We implement sum operation for spatial attention module and channel attention module. The ensemble refers to simple ensemble strategy: average of different manipulated denoised images. The input of networks are joint manipulation types which consists of spatial domain manipulations and frequency domain manipulations. The results are shown on Table \ref{table:den_psnr_results_comparison}. The sample visual results of the base denoisers and our fusion models are shown on Fig \ref{fig:dncnn-al}.

\begin{table*}[t]
	\centering
	\setulcolor{red}
	\setul{0.25ex}{0.25ex}
	\resizebox{\textwidth}{40mm}{
	\begin{tabular}{ccccccc}
		\toprule
		\makecell[c]{Backbone\\denoiser}
		& \makecell[c]{Noise\\level} 
		& \makecell[c]{Baseline\\results} 
		& \makecell[c]{Ensemble\\(SM - FM - Joint)}
		& \makecell[c]{Spatial attention\\(SM - FM - Joint)}
		& \makecell[c]{Channel attention\\(SM - FM - Joint)} 
		& \makecell[c]{Ours full\\(SM - FM - Joint)} \\ \cline{1-7}
		\noalign{\smallskip}
		
		\multirow{5}{*}{DnCNN~\cite{zhang2017beyond}}
		& 10 & \cellcolor{gray} 33.30 & 33.38 32.29 33.11 & 33.48 33.52 \ul{33.56} & 33.39 33.37 \ul{33.45} & 33.55 33.52 \textbf{33.58} \\
		& 20 & \cellcolor{gray} 29.72 & 29.78 29.25 29.67 & 29.84 \ul{29.92} 29.90 & 29.78 29.73 \ul{29.79} & 29.99 29.98 \textbf{30.03} \\
		& 30 & \cellcolor{gray} 27.68 & 27.74 27.34 27.66 & 27.83 \ul{28.02} \ul{28.02} & 27.74 27.70 \ul{27.75} & 28.12 28.10 \textbf{28.16} \\
		& 40 & \cellcolor{gray} 26.19 & 26.24 25.87 26.18 & 26.41 \ul{26.72} 26.70 & 26.24 26.25 \ul{26.29} & 26.88 26.89 \textbf{26.91} \\
		& 50 & \cellcolor{gray} 24.96 & 25.01 24.67 24.96 & 25.26 \ul{25.62} 25.55 & 25.01 25.13 \ul{25.15} & 25.96 25.97 \textbf{25.99} \\
		\noalign{\smallskip}
		
		\multirow{5}{*}{MemNet~\cite{tai2017memnet}} 
		& 10 & \cellcolor{gray} 33.40 & 33.52 32.36 33.25 & 33.52 33.55 \ul{33.60} & 33.52 33.43 \ul{33.54} & 33.64 33.52 \textbf{33.65} \\
		& 20 & \cellcolor{gray} 29.71 & 29.79 29.10 29.63 & 29.85 29.94 \ul{29.99} & 29.79 29.78 \ul{29.84} & 30.05 29.95 \textbf{30.06} \\
		& 30 & \cellcolor{gray} 27.61 & 27.68 27.10 27.55 & 27.83 28.06 \ul{28.07} & 27.68 27.77 \ul{27.81} & 28.16 28.14 \textbf{28.18} \\
		& 40 & \cellcolor{gray} 26.11 & 26.17 25.64 26.06 & 26.36 26.70 \ul{26.78} & 26.17 26.37 \ul{26.39} & 26.92 26.93 \textbf{26.94} \\
		& 50 & \cellcolor{gray} 24.94 & 24.99 24.51 24.92 & 25.22 25.62 \ul{25.80} & 24.99 \ul{25.29} 25.27 & 25.92 26.01 \textbf{26.02} \\  
		\noalign{\smallskip}
		
		\multirow{5}{*}{RIDNet~\cite{anwar2019real}}
		& 10 & \cellcolor{gray} 33.58 & 33.67 32.41 33.35 & 33.66 33.65 \ul{33.70} & \ul{33.67} 33.59 \ul{33.67} & \textbf{33.73} 33.63 \textbf{33.73} \\
		& 20 & \cellcolor{gray} 29.86 & 29.91 29.17 29.73 & 29.93 29.98 \ul{30.06} & 29.91 29.89 \ul{29.93} & 30.10 30.06 \textbf{30.11} \\
		& 30 & \cellcolor{gray} 27.71 & 27.76 27.13 27.61 & 27.87 \ul{28.11} \ul{28.11} & 27.76 27.83 \ul{27.87} & 28.22 28.19 \textbf{28.24} \\
		& 40 & \cellcolor{gray} 26.13 & 26.18 25.65 26.07 & 26.35 26.81 \ul{26.85} & 26.18 26.42 \ul{26.44} & 26.97 26.97 \textbf{27.01} \\
		& 50 & \cellcolor{gray} 24.90 & 24.95 24.50 24.88 & 25.17 \ul{25.60} 25.55 & 24.95 \ul{25.32} \ul{25.32} & 26.01 26.06 \textbf{26.08} \\
		
		\bottomrule 
	\end{tabular}}
	\caption{Gaussian denoising PSNR ($dB$) results of the baseline networks, the averaging ensemble, our spatial attention module, our channel attention module, and our fusion model. We include the ablations using only our spatially-manipulated (SM) or frequency-manipulated (FM) images, rather than all (Joint) manipulations. Best results in bold, best per attention mechanism are underlined.}
	\vspace{-0.15cm}

	\label{table:den_psnr_results_comparison}
\end{table*}

\subsection{Discussion}
\textbf{Results analysis:} According to the PSNR(dB) shown on Table \ref{table:den_psnr_results_comparison}, we can find that our fusion models achieve good performance in denoising compared with baseline models. The channel attention modules which assign weights to different manipulated images perform better than simple average method. The spatial attention modules get similar results with fusion models at low noise levels, but the fusion models achieve better results at high noise levels than just using spatial attention modules. 

\textbf{Other relevant attempts:} In addition to the experiments shown above, we also train our fusion models based on other kinds of pre-trained denoisers: DnCNN with SFM, MemNet with SFM, RIDNet with SFM. Also, we explore the effect of clip to the pre-trained denoisers and use our fusion models on them again. Otherwise, we use DC coefficients instead of pixel values into deep neural networks, but it shows worse performance than corresponding results which use pixel values as input.

\section{Retrain DnCNN with Auxiliary Loss}
\subsection{Results}
At this experiment, we use training data-set BSD400 and test data-set BSD68 to implement three configurations and other parameters setting are as the same as the pre-trained denoisers shown above.
\begin{itemize}
    \setlength{\itemsep}{0pt}
    \setlength{\parsep}{0pt}
    \setlength{\parskip}{0pt}
    \item retrain DnCNN with auxiliary loss from $L_{1}$ error estimator.
    \item retrain DnCNN with auxiliary loss from $L_{2}$ error estimator.
    \item retrain DnCNN with auxiliary loss from DnCNN learning images.
\end{itemize}

\begin{figure}
    \centering
\subfloat[$\sigma=10$ \label{fig:f1}]{
    \includegraphics[width=0.3\textwidth]{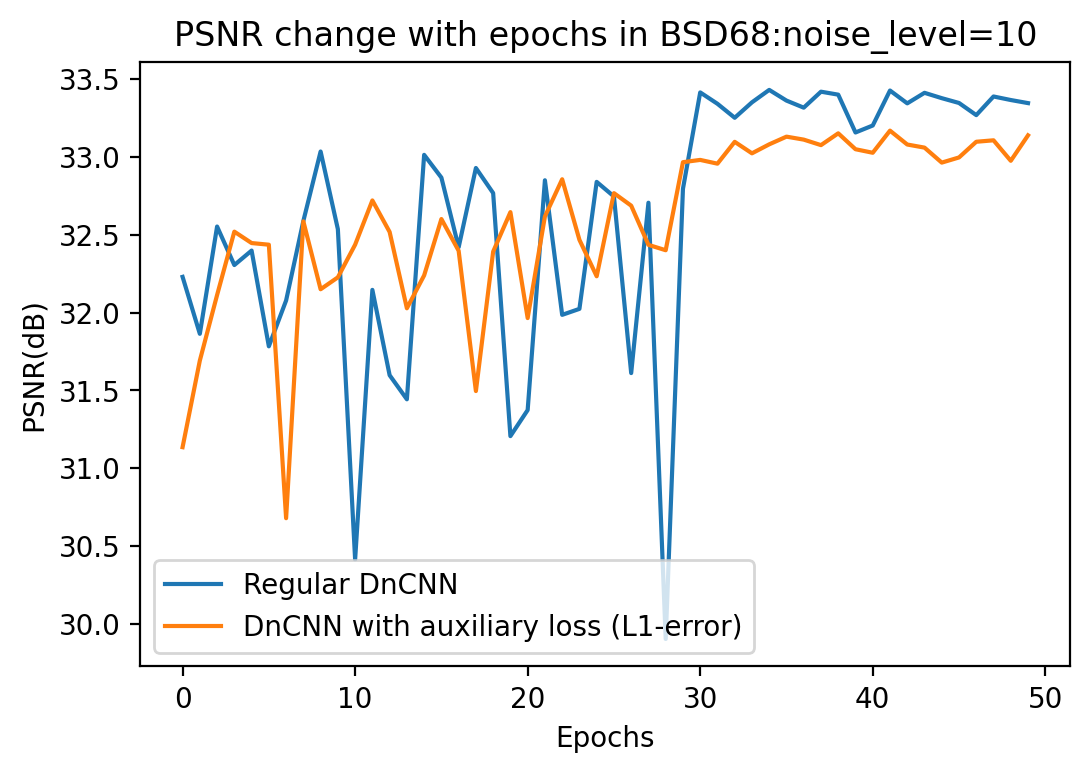}}
\subfloat[$\sigma=30$ \label{fig:f2}]{
    \includegraphics[width=0.3\textwidth]{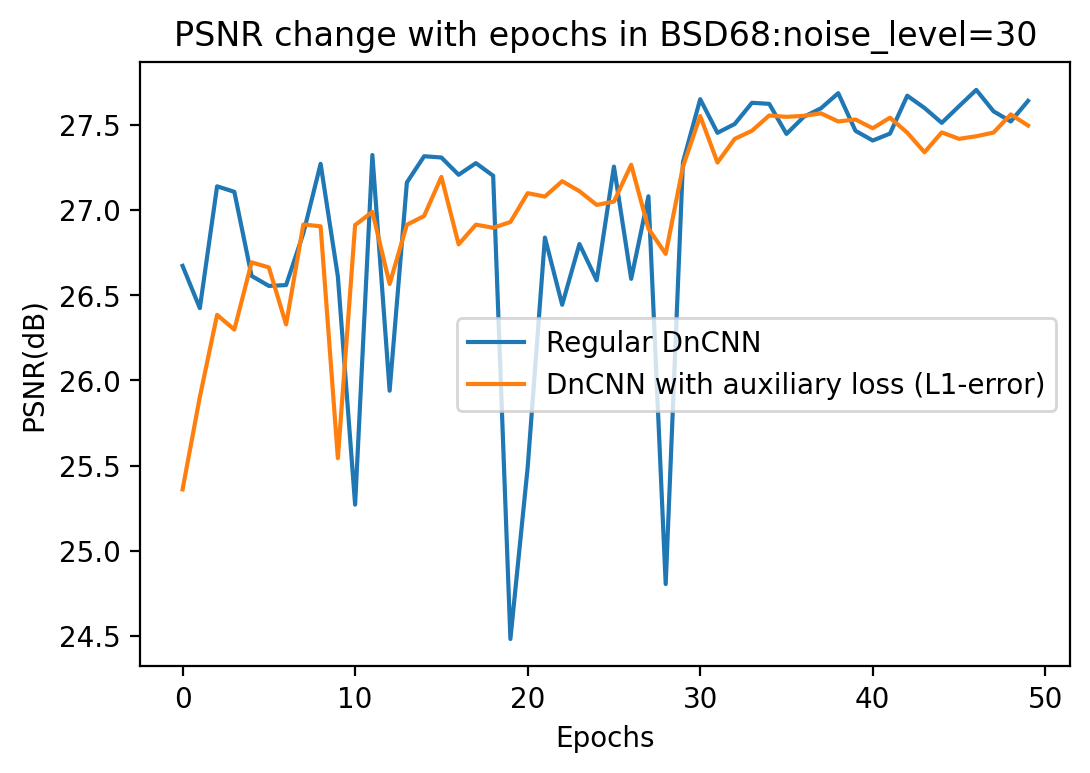}}
\subfloat[$\sigma=50$\label{fig:f3}]{
    \includegraphics[width=0.3\textwidth]{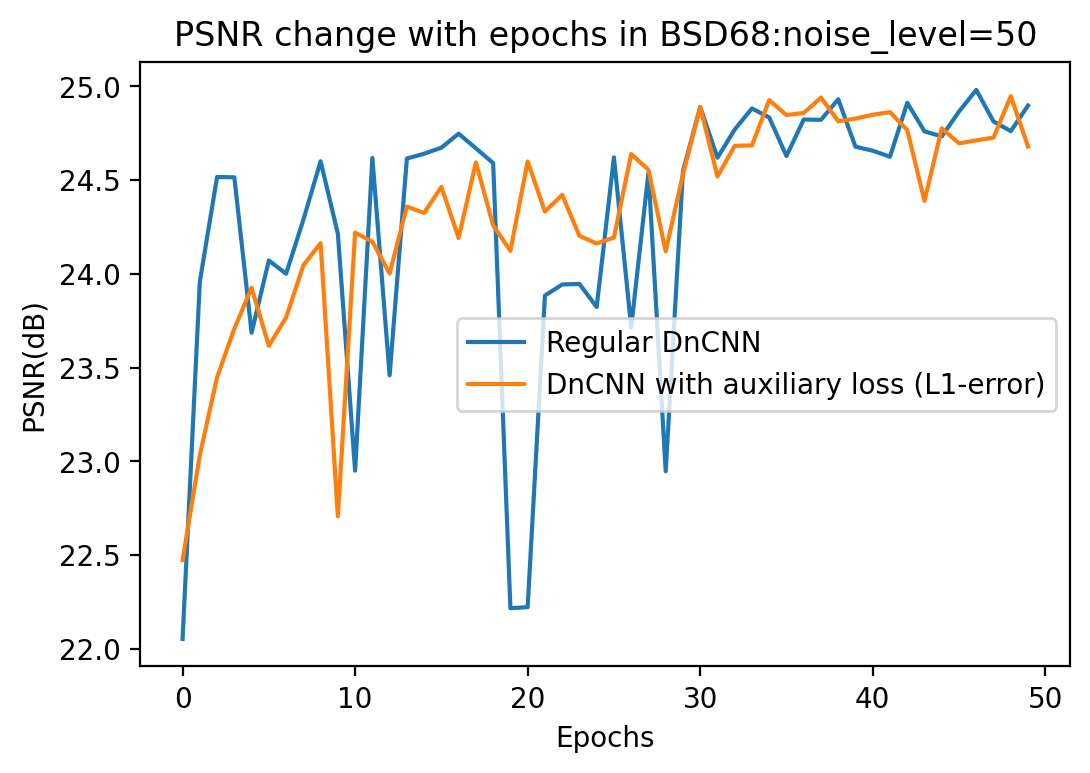}}
\caption{PSNR(dB) change along with epochs for regular DnCNN and DnCNN with with auxiliary loss from $L_{1}$ error estimator.}
\label{fig:dncnn-alg}
    \end{figure}

Firstly, we try to compare the PSNR results between three configurations and the results of regular DnCNN, we find that there is no significant difference for DnCNN with auxiliary loss from $L_{1}$ error estimator and DnCNN with auxiliary loss from DnCNN learning images. However, DnCNN with auxiliary loss from $L_{2}$ error estimator shows worse performance than regular DnCNN, which means the auxiliary loss is 
counterproductive to the main denoising task. Then, we observe the PSNR change along with epochs and find that PSNR change of DnCNN with auxiliary loss from $L_{1}$ error estimator is more stable than regular DnCNN, which means the oscillation is alleviated by usage of auxiliary loss. The figure\ref{fig:dncnn-alg} shows the PSNR change of two models at noise level 10, 30 and 50.

\subsection{Discussion}
From our experiments, we find our designs do not make progress in denoising performance compared with regular models, but oscillation is alleviated by taking the $L_{1}$ error estimator as auxiliary tasks. We also tests other methods such as ensemble of denoised image and output of error estimator, but it also did not achieve significant progress.

\chapter{Conclusion \&Future work}
\section{Conclusion}
In this project, we focus on exploring the uncertainty existing in the denoisers based on convolutional neural networks. From the point of view of ensemble learning, we implemented different manipulations to noisy images from the perspective of spatial domain and frequency domain and proposed decouple dual attention fusion models to minimize the uncertainty of images denoised by pre-traind denoisers. The results show that our methods make good progress in image denoising performance. Additionally, we also tested other methods such as auxiliary loss to improve the denoising networks, although the results show that this approach requires further development.

\section{Future work}
In the future, we can test our models in more data-sets such as real images and improve the fusion method of two attention modules to achieve better performance. Otherwise, we can find suitable methods to extend our model to other image processing tasks. On the other hand, how to make better use of error estimator, thus further minimize the uncertainty of denoising networks still needs more attentions.

\clearpage
\pagestyle{numberonly}
\printbibliography

\end{document}